\documentclass{article}
\usepackage{cite,tikz,amsmath,amssymb,graphicx,hyperref,booktabs}
\usepackage{comment,multirow,makecell,booktabs}
\usepackage{algorithmic}
\usepackage{color}
\usepackage{arydshln}
\usepackage{siunitx}
\usepackage{textcomp}
\usepackage{xcolor}
\usepackage{hyperref}

\usepackage[preprint]{spconf}
\copyrightnotice{%
\parbox{\textwidth}{\footnotesize
© 2026 IEEE. Personal use of this material is permitted. Permission from IEEE must be obtained for all other uses, in any current or future media, including reprinting/republishing this material for advertising or promotional purposes, creating new collective works, for resale or redistribution to servers or lists, or reuse of any copyrighted component of this work in other works.
}}

\title{CALM: JOINT CONTEXTUAL ACOUSTIC-LINGUISTIC MODELING FOR \\PERSONALIZATION OF MULTI-SPEAKER ASR}
\name{Muhammad Shakeel$^1$, Yosuke Fukumoto$^1$, Chikara Maeda$^1$, Chyi-Jiunn Lin$^2$, Shinji Watanabe$^2$}
\address{
  $^1$Honda Research Institute Japan Co., Ltd., Saitama, Japan\\
  $^2$Carnegie Mellon University, Pittsburgh, PA, USA}
\begin{document}
\ninept
\maketitle
\begin{abstract}
We present \textbf{CALM}, a joint \textbf{C}ontextual \textbf{A}coustic-\textbf{L}inguistic \textbf{M}odeling framework for multi-speaker automatic speech recognition (ASR). In personalized AI scenarios, the joint availability of acoustic and linguistic cues naturally motivates the integration of target-speaker conditioning with contextual biasing in overlapping conversations.
CALM implements this integration in an end-to-end framework through speaker embedding–driven target-speaker extraction and dynamic vocabulary–based contextual biasing. We evaluate CALM on simulated English (LibriSpeechMix) and Japanese (Corpus of Spontaneous Japanese mixtures, CSJMix). On two-speaker mixtures, CALM reduces biased word error rate (B-WER) from 12.7 to 4.7 on LibriSpeech2Mix and biased character error rate (B-CER) from 16.6 to 8.4 on CSJMix2 (eval3), demonstrating the effectiveness of joint acoustic-linguistic modeling across languages. We additionally report results on the AMI corpus (IHM-mix condition) to validate performance on standardized speech mixtures.
\end{abstract}
\section{INTRODUCTION}
\label{sec:intro}
Single-speaker automatic speech recognition (ASR) systems have achieved state-of-the-art (SOTA) performance across many speech-processing tasks \cite{WavLM,whisper,peng25c_interspeech}. However, in multi-speaker settings \cite{UME} with overlapping speech \cite{ShinjiCHiME6,M2Met,Notsofar1} and conversation-specific vocabulary \cite{le21_interspeech,biasphraseboosted,shakeel24_interspeech}, performance degrades substantially, limiting personalization in real-world applications \cite{meeting-recognition,nttmultispeakerasrdasr} (e.g., meeting minutes and group discussions) that require accurate, speaker-attributed transcripts. The degradation is twofold, reflecting (i) acoustic challenges—suboptimal attribution under overlap due to interference from non-target talkers; and (ii) linguistic challenges—limited adaptability to domain-specific language, including proper nouns, acronyms, and technical jargons that are sparse or unseen in training.

Two complementary families of approaches have emerged to address the challenges of conversational ASR frameworks. Acoustic errors are commonly mitigated through end-to-end (E2E) modeling that optimize speaker attribution (i.e., determining who is speaking). Typical methods condition ASR on target-speaker (TS) embeddings \cite{delcroix19_interspeech,location-guided-TSE,moriya22_interspeech,zili1,ConformerTSASR,masumura23_interspeech}, token-level speaker labels \cite{kanda21b_interspeech,kanda22b_interspeech}, voice activity detection (VAD) \cite{medennikov20_interspeech,onlineNeuralSpeaker,maeda25_interspeech}, or diarization-aware \cite{DiCoW,TSASR-whisper} cues to extract speaker-discriminative information in overlapping speech, while recent work leverages stronger speech encoders \cite{WavLM,whisper,peng25c_interspeech} to further improve TS ASR acoustic robustness \cite{SQ-Whisper,meng24c_interspeech}. In contrast, linguistic adaptability are typically handled by two approaches; shallow fusion \cite{zhao19d_interspeech}, where an external language model is integrated with the E2E decoder during inference, and contextual biasing (CB) methods \cite{le21_interspeech,acoustic-and-semantic} that employ context encoders, lightweight adapter modules \cite{contextual-adapters}, or architectural extensions with phrase-level prediction networks \cite{huang23d_interspeech} and intermediate contextual biasing networks \cite{GuidedAttention,shakeel24_interspeech}. Recent advances further explore dynamic vocabulary expansion \cite{copyne,dynamicvocab} for handling out-of-vocabulary terms and serialized output training (SOT) \cite{CMT-LLM}, often augmented with large language model (LLM) prompting \cite{llmcontextualasr}, to introduce contextual information directly into the decoding process.

Despite these advances, prior work has primarily been evaluated under single-speaker \cite{biasphraseboosted} or non-overlapping conditions, or limited to SOT-based frameworks \cite{CMT-LLM}, which often degrade in highly overlapping or variable multi-speaker scenarios. In these conditions, scalability and effective integration with speaker-specific cues remain challenging, thereby neglecting the crucial inter-dependencies between acoustic and linguistic contexts. Moreover, addressing acoustic errors alone has proven insufficient in multi-speaker scenarios, as linguistic errors—particularly those arising from unseen vocabularies or spontaneous conversational structures—remain orthogonal sources of degradation. These limitations motivate the development of a joint acoustic-linguistic framework that leverages speaker-aware acoustic representations to inform linguistic adaptation, enabling personalized and robust multi-speaker ASR.

In this work, we propose CALM, a joint Contextual Acoustic-Linguistic Modeling framework for multi-speaker automatic speech recognition (ASR). CALM implements this integration in an end-to-end framework through speaker embedding–driven target-speaker extraction and dynamic vocabulary–based contextual biasing, ensuring robust contextual adaptation in overlapping speech. We evaluate CALM on English LibriSpeechMix mixtures \cite{kanda20b_interspeech}, AMI \cite{AMI} and the corpus of spontaneous Japanese (CSJ) \cite{csj}. Experiments show substantial gains in word error rate (WER), character error rate (CER), unbiased WER/CER (U-WER/U-CER), and biased WER/CER (B-WER/B-CER) compared to standalone TS ASR and CB baselines, confirming the effectiveness of jointly leveraging linguistic and acoustic adaptation. Our contributions are threefold: (i) a unified acoustic-linguistic framework that integrates TS conditioning with dynamic vocabulary expansion (ii) a scalable mechanism for personalized contextual adaptation, effective across varying biasing list sizes and dynamic multi-speaker conversational scenarios; and (iii) extensive evaluation on LibriSpeechMix, AMI and CSJ datasets, demonstrating substantial gains over state-of-the-art baselines.
\vspace{-10px}
\section{CALM}
\label{sec:calm}
\vspace{-5px}
We consider a multi-speaker conversation scenario where the input mixture is represented as \begin{math} X = \sum_{c=1}^C Y^c \odot S^c + G, \quad X \in \mathbb{R}^{T},\end{math} with clean sources \(S^c\) of $C$ speakers, activity masks \(Y^c \in [0,1]^T\), and additive noise \(G\). In this work, we employ \texttt{WavLM-Large} \cite{WavLM} as an upstream model to extract frame-level features \begin{math} X^{\text{fe}} \in \mathbb{R}^{T^{\text{fe}} \times D}.\end{math} These are projected into the encoder space using a learned linear transformation \(Z^{\text{fe}} = W^{\text{proj}} X^{\text{fe}}, \quad Z^{\text{fe}} \in \mathbb{R}^{T^{\text{fe}} \times D^{\text{fe}}}\).
Since this is a personalization scenario, we can expect to obtain enrollment utterances as target-speaker cues $(C_s \in \mathbb{R}^{T_\text{s} \times F_\text{s}}$). We employ ECAPA-TDNN \cite{ecapa} with attentive pooling, followed by a RawNet3 projector as a speaker encoder (SpeakerEnc) to obtain a speaker embedding $E_{\text{s}}$ of size $D^{\text{emb}}$ and represented as:
\vspace{-5px}
\begin{equation}
E_s = \mathrm{SpeakerEnc}(C_s) \in \mathbb{R}^{D^{\text{emb}}}.
\label{eq:emb}
\end{equation}
We use Conformer \cite{gulati20_interspeech} as an audio encoder (AudioEnc) comprising of $L$ encoder layers to map the projected features to its input and is given by:
\vspace{-5px}
\begin{equation}
H_{(l)} = \mathrm{AudioEnc}_{(l)}(Z^{\text{fe}}).
\label{eq:audio_enc}
\end{equation}
The embedding from \eqref{eq:emb} is then employed as a speaker adaptation mechanism using a FiLM-based \cite{FiLM} modulation of the hidden representations of the intermediate and final audio encoder states and is represented as:
\vspace{-10px}
\begin{equation}
\hat{H}_{(l)} = \gamma(E_s) \odot H_{(l)} + \beta(E_s),
\label{eq:film_modulation}
\end{equation}
where \(\gamma(E_s), \beta(E_s) \in \mathbb{R}^{D^{\text{enc}}}\), \(\hat{H}_{(l)} \in \mathbb{R}^{T^{\text{enc}} \times D^{\text{enc}}}\) and \( T^{\text{enc}} < T^{\text{fe}}  \).

Since we want to achieve robust contextual adaptation in overlapping speech, we use Transformer \cite{transformer} with mean pooling following the setup in \cite{sudo25_interspeech} as a dynamic vocabulary-based contextual biasing encoder (BiasEnc) to encode a biasing list \(B=\{b_1, \ldots, b_N\}\) into phrase-level vector representations \(V=[v_1, \ldots, v_N]\), where each biasing phrase in $B$ is incorporated as a single dynamic token to build a dynamic vocabulary \(\mathcal{V}^{\text{d-vocab}}=\{\langle b_1\rangle, \ldots, \langle b_N\rangle\}\) of size $N$ and is given by:
\vspace{-5px}
\begin{equation}
V = \mathrm{BiasEnc}(B), \quad V \in \mathbb{R}^{N\times{D^{\text{bias}}}}.
\label{eq:phrases}
\end{equation}
Similar to \cite{sudo25_interspeech}, we formulate the dynamic vocabulary-based intermediate and final output layers by extending the static vocabulary \(\mathcal{V}^{\text{stat}}\) of size $M$ with the dynamic vocabulary \(\mathcal{V}^{\text{d-vocab}}\) of size $N$ as $V$ and is given by:
\vspace{-5px}
\begin{align}
    \label{eq:s_score}
    & O^{\text{stat}}_{(l)} = \mathrm{Linear}(\hat{H}_{(l)}), \quad O^{\text{stat}}_{(l)} \in \mathbb{R}^{T \times M},\\
    \label{eq:d_score}
    & O^{\text{d-vocab}}_{(l)} = \frac{\mathrm{Linear}(\hat{H}_{(l)}) \mathrm{Linear}(V^\top)}{\sqrt{D^{\text{bias}}}}, \quad  O^{\text{d-vocab}}_{(l)} \in \mathbb{R}^{T \times N}, \\
    \label{eq:prob}
    & O_{(l)} = \mathrm{Softmax}(\mathrm{concat}(O^{\text{stat}}_{(l)}, O^{\text{d-vocab}}_{(l)})),  O_{(L)} \in \mathbb{R}^{T \times (M+N)}
\end{align}

where $O^{\text{stat}}_{(l)}$ and $O^{\text{d-vocab}}_{(l)}$ denote the alignment scores for static and dynamic vocabulary, respectively. In \eqref{eq:prob} we employ weighted softmax adapted from \cite{dynamicvocab} to avoid over/under-biasing during inference and control the dynamic token probabilities with a biasing weight ($\mu$) as a hyperparameter. Finally we optimize the model parameters and compute the CTC loss ($\mathcal{L}_{\text{ctc}}$) and interCTC loss ($\mathcal{L}_{\text{interctc}}$) by minimizing the negative log-likelihood. Given the combined output distribution $O_{(l)}$ from \eqref{eq:prob}, the losses are defined as:
\vspace{-5px}
\begin{equation}
    \mathcal{L}_{\text{ctc}} = - \log P(Y \mid \hat{H}_{(L)}, B) = - \log P(Y \mid O_{(L)}).
\label{eq:loss_ctc}
\end{equation}
\vspace{-10px}
\begin{equation}
    \mathcal{L}_{\text{interctc}} = -\frac{1}{|L|} \sum_{l \in L} \log P(Y \mid O_{(l)}).
\label{eq:inter_ctc}
\end{equation}
Following \cite{sudo25_interspeech}, we also employ CTC self-conditioning, where the posterior probabilities $O_{(l)}$ from intermediate layers are  added back to the encoder input at subsequent layers, enabling the dynamic  vocabulary information to propagate throughout the encoder.

In addition to CTC, we use a dynamic vocabulary-based attention decoder \cite{dynamicvocab} that predicts dynamic tokens autoregressively. Given audio encoder states $\hat{H}_{(L)}$ and biasing representations $V$, we compute the attention loss ($\mathcal{L}_{\text{att}}$) as the negative log-likelihood and is given by:
\vspace{-10px}
\begin{equation}
\mathcal{L}_{\text{att}} = - \log P(Y \mid \hat{H}_{(L)}, B).
\end{equation}
To regularize acoustic modeling, we attach a VAD head following \cite{maeda25_interspeech} to the final adapted encoder states $\hat{H}_{(L)}$. Frame-level target-speaker activity posteriors are computed as:
\vspace{-5px}
\begin{equation}
P^{\text{vad}} = \sigma(W^{\text{vad}} \hat{H}_{(L)} + b^{\text{vad}}),
\vspace{-5px}
\end{equation}
with $P^{\text{vad}} \in [0,1]^{T^{\text{enc}}}$.  
The VAD loss ($\mathcal{L}_{\text{vad}}$) is then the binary cross-entropy between predictions and ground-truth activity labels $Y^{\text{vad}}$:
\vspace{-5px}
\begin{equation}
\mathcal{L}_{\text{vad}} = \mathrm{BCE}(P^{\text{vad}}, Y^{\text{vad}}).
\end{equation}
The final training objective is a weighted multitask loss that combines CTC, attention, interCTC, and VAD losses:
\vspace{-5px}
\begin{equation}
\mathcal{L}_{\text{total}}
= \lambda_{\text{ctc}} \, \mathcal{L}_{\text{ctc}} 
+ \lambda_{\text{vad}} \, \mathcal{L}_{\text{vad}}
+ (1 - \lambda_{\text{ctc}} - \lambda_{\text{vad}})\, \mathcal{L}_{\text{att}}.
\vspace{-5px}
\end{equation}
Here, the CTC itself optimizes both intermediate and last layers:
\vspace{-5px}
\begin{equation}
\mathcal{L}_{\text{ctc}}
= (1 - \lambda_{\text{interctc}})\,\mathcal{L}_{\text{ctc}}^{(L)}
+ \lambda_{\text{interctc}} \,\mathcal{L}_{\text{interctc}},
\vspace{-5px}
\end{equation}
where $\mathcal{L}_{\text{ctc}}^{(L)}$ is computed at the last layer $L$, $\mathcal{L}_{\text{interctc}}$ from selected intermediate layers and $\lambda_{\{\cdot\}}$ are tunable weights.

\begin{figure}[t]
\begin{center}
{\resizebox{7cm}{!}{\includegraphics[width=\textwidth]{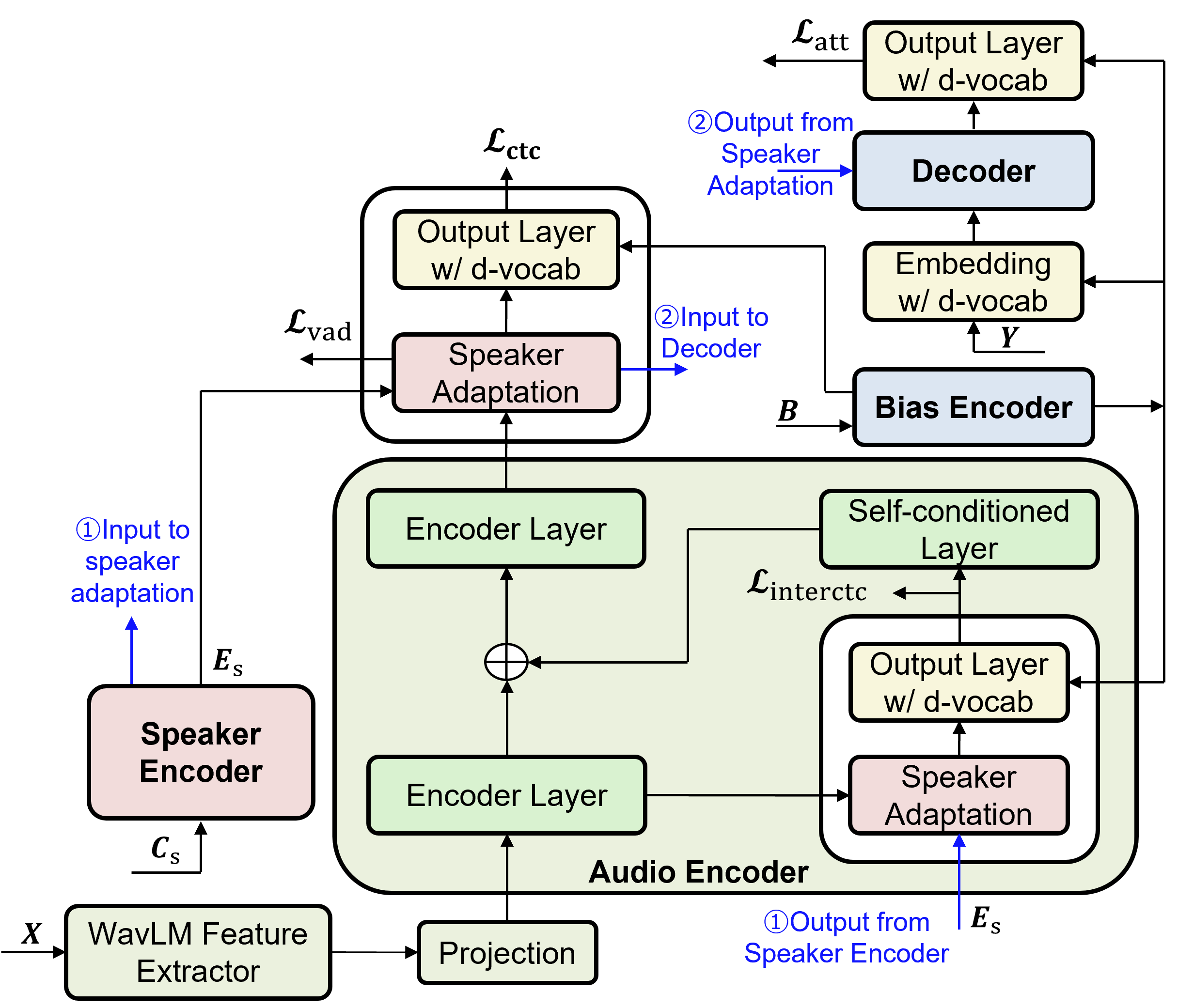}}}
\caption{Illustration of CALM framework.}
\label{fig:CALM} 
\end{center}
\vspace{-25px}
\end{figure}
\begin{table*}[t]
\centering
\caption{WER (U-WER/B-WER) results on LibriSpeechMix datasets with different biasing list sizes. A3 and A4 employ biasing weight ($\mu{=}0.1$) at the inference stage. Best results are bolded.}
\label{tab:librispeech_main}
\setlength{\tabcolsep}{4pt}
\renewcommand{\arraystretch}{1.0}
\resizebox{\textwidth}{!}{%
\begin{tabular}{@{}c l  cccc  cccc@{}}
\toprule
 &  & \multicolumn{4}{c}{\textbf{LibriSpeech2Mix}} & \multicolumn{4}{c}{\textbf{LibriSpeech3Mix}} \\
\cmidrule(lr){3-6}\cmidrule(lr){7-10}
\textbf{ID} & \textbf{Method} &
$N{=}0$ & $N{=}100\times2$ & $N{=}500\times2$ & $N{=}1000\times2$ &
$N{=}0$ & $N{=}100\times3$ & $N{=}500\times3$ & $N{=}1000\times3$ \\
\cmidrule(lr){1-2}\cmidrule(lr){3-6}\cmidrule(lr){7-10}
\multicolumn{10}{l}{\textbf{Baselines (reproduced)} \cite{maeda25_interspeech}}\\
\quad A1 & \makecell[l]{TS-ASR w/ \\ (SC-CTC/ATTN)} &
4.6 (3.6/12.9) & 4.6 (3.6/12.9) & 4.6 (3.6/12.9) & 4.6 (3.6/12.9) &
9.9 (9.0/17.2) & 9.9 (9.0/17.2) & 9.9 (9.0/17.2) & 9.9 (9.0/17.2) \\
\quad A2 &  \makecell[l]{A1 w/ VAD loss} &
\textbf{4.3} (\textbf{3.3}/\textbf{12.7}) &
4.3 (\textbf{3.3}/12.7) &
4.3 (\textbf{3.3}/12.7) &
4.3 (\textbf{3.3}/12.7) &
\textbf{9.2} (\textbf{8.3}/\textbf{17.0}) &
9.2 (\textbf{8.3}/17.0) &
9.2 (\textbf{8.3}/17.0) &
9.2 (\textbf{8.3}/17.0) \\
\cmidrule(lr){1-2}\cmidrule(lr){3-6}\cmidrule(lr){7-10}
\multicolumn{10}{l}
{\textbf{Proposed}}\\
\quad A3 & \makecell[l]{A1 w/ \\dynamic vocab.} &
5.3 (4.1/14.3) &
3.9 (3.8/4.3) &
4.0 (4.0/4.4) &
4.2 (4.1/\textbf{4.7}) &
11.1 (10.2/19.0) &
9.2 (9.4/7.8) &
9.4 (9.6/8.0) &
9.6 (9.7/8.7) \\
\quad A4 & \makecell[l]{A3 w/ \\ VAD loss} &
4.9 (3.7/14.7) &
\textbf{3.6} (3.5/\textbf{4.1}) &
\textbf{3.7} (3.7/\textbf{4.1}) &
\textbf{4.1} (4.0/4.9) &
10.2 (9.2/18.7) &
\textbf{8.4} (8.5/\textbf{6.9}) &
\textbf{8.8} (8.9/\textbf{7.6}) &
\textbf{9.1} (9.3/\textbf{8.3}) \\
\bottomrule
\end{tabular}
}
\vspace{-10px}
\end{table*}
\section{Experiments}
\label{sec:experiment}
The CALM framework is built on ESPnet \cite{espnet}, pairing a Conformer encoder with a Transformer decoder. The Conformer has 12 layers with 4 heads and 1024 linear units (kernel size 31) and applies self-conditioned interCTC similar to \cite{sudo25_interspeech} at layers 3, 6, and 9; the decoder is a 6-layer Transformer with 4 heads and 2048 units. For the input stack, we use \texttt{conv2d} on LibriSpeechMix and AMI, and \texttt{conv2d6} on CSJMix. We inject linguistic context using a compact 6-layer Transformer-based bias encoder with 4 heads and 1024 units. Speaker information comes from an ECAPA-TDNN \cite{ecapa} encoder (scale 8; 1024 linear units; 1536-dim output), passed through a RawNet3 projector (192-dim) and adapted with FiLM. The speaker frontend uses MelSpec Torch (nfft:512, window:400, hop:160, 80 mel banks). Audio is extracted with frozen WavLM-Large features, and we apply SpecAugment across all corpora. During training, we randomly construct a biasing list $B$ per batch, yielding $N{=}50{\sim}200$ phrases following \cite{sudo25_interspeech}, and train the framework with a static vocabulary size $M{=}5000$. Enrollment procedures differ by corpus. Training follows joint CTC/attention with interCTC and an auxiliary VAD loss (weights: CTC 0.3, interCTC 0.5, VAD 0.15; self-conditioning enabled) and uses Adam with a maximum learning rate of $2\times10^{-3}$. Schedules are: LibriSpeechMix—15k warmup, 36M batch bins and 70 epochs; CSJMix—25k warmup, 40M batch bins and 50 epochs; AMI—15k warmup, 40M batch bins and 30 epochs.

\noindent\textbf{Datasets.} We evaluate CALM on LibriSpeechMix, CSJMix, and AMI corpora. LibriSpeechMix \cite{kanda20b_interspeech} extends LibriSpeech \cite{librispeech} with WHAM! noise \cite{wham}, creating 960 h of two- and three-speaker mixtures with random delays between overlapping speakers \cite{maeda25_interspeech}. CSJMix applies the same procedure to the CSJ corpus (581 h), enabling cross-linguistic evaluation on Japanese. Both use clean mixtures for testing. We adopt the AMI standardized mixtures (IHM-Mix condition) with 4–5 speakers (79.4 h) \cite{zili1}. Speaker enrollment is required: 5s per speaker for LibriSpeechMix/CSJMix \cite{maeda25_interspeech}, and 15s for AMI \cite{zili1}, where unlike prior work, enrollment segments are taken from the same meeting IDs to better match real conditions.

\noindent\textbf{Biasing list creation.}\footnote{The link will appear in the camera-ready version of this paper.}
We follow the biasing list construction of \cite{le21_interspeech} and extend it to multi-speaker mixtures. For LibriSpeechMix and CSJMix, biasing lists are built per speaker and concatenated to size $N \times C$ (see Section \ref{sec:calm}). In LibriSpeechMix, rare words are defined as those outside the 5,000 most frequent words in the training set ($\approx$ 90\% coverage), while in CSJMix (character-based), rare characters are outside the 1,500 most frequent common characters. For each utterance, reference words/characters outside the common set are included as biasing words/characters, and distractors are randomly sampled from the rare set to reach size $N$. For AMI, lists are created per utterance rather than per speaker: words outside the frequent 1,000 are treated as rare, with reference rare words included and distractors sampled to the target size $N$. Unlike the simulated corpora, AMI uses a single list per utterance spanning all active speakers.

\noindent\textbf{Evaluation Metrics.} 
We evaluate CALM framework following the metrics in \cite{le21_interspeech}. WER/CER is calculated on all words or characters, U-WER/U-CER on words/characters outside the biasing list, and B-WER/B-CER on words/characters inside the biasing list. Insertion errors are assigned to B-WER/B-CER if the inserted word or character appears in the biasing list, and to U-WER/U-CER otherwise. 
\vspace{-7px}
\subsection{Main results}
Table \ref{tab:librispeech_main} presents results on LibriSpeech2Mix and LibriSpeech3Mix with varying biasing list sizes. Among all variants, A4 achieves the most balanced performance, consistently improving WER, U-WER, and B-WER across both two- and three-speaker mixtures. Baseline TS-ASR models (A1/A2) achieve competitive WER and U-WER, but their B-WER remains high (12.7 points on LibriSpeech2Mix, 17.0 points on LibriSpeech3Mix), reflecting the limits of acoustic modeling without linguistic adaptation. In contrast, the proposed CALM framework (A3/A4) achieves substantial reductions in B-WER, particularly with larger biasing lists. On LibriSpeech2Mix with an effective biasing list of 2000 entries, B-WER is reduced from 12.7 to 4.7 absolute points, while on LibriSpeech3Mix with 3000 entries, it is reduced from 17.0 to 8.3 absolute points. Notably, these gains are accompanied by consistent reductions in WER and U-WER, demonstrating that the improvements extend beyond biasing words and enhance overall recognition robustness. Unlike conventional contextual biasing approaches that insert static tokens only at intermediate layers \cite{shakeel24_interspeech} or exclusively at the last layer \cite{biasphraseboosted}—which often degrades B-WER as list size increases—our framework adapts the dynamic vocabulary expansion \cite{sudo25_interspeech} design where static and dynamic tokens are jointly expanded into intermediate self-conditioned CTC layers, providing the encoder with richer lexical information and balanced probability allocation across token types. This prevents dynamic tokens from overbaising the entire network and ensures that distractor probability distributions are equally regularized throughout the encoder. Consequently, both B-WER and U-WER improve as the biasing list grows. The speaker adaptation process further enhances this effect by selectively amplifying biasing phrases relevant to the enrolled speaker, while the biasing weight ($\mu$) (see Table \ref{tab:biasingweight}) applied at inference provides balanced performance. These results confirm that combining dynamic vocabulary expansion with speaker-aware adaptation and auxiliary VAD regularization yields an effective configuration for robust multi-speaker ASR.
\vspace{-5px}
\subsection{Effect of biasing weight}
Following \cite{sudo25_interspeech}, we apply a biasing weight ($\mu$) (see Section \ref{sec:calm}) during inference to balance dynamic and static vocabularies in the weighted softmax \cite{dynamicvocab}. Larger values of $\mu$ overweight biasing tokens, improving B-WER but often degrading overall recognition, while smaller values maintain more stable performance. As shown in Table~\ref{tab:biasingweight}, $\mu{=}0.1$ achieves the best trade-off (WER 4.1, B-WER 4.9 on LibriSpeech2Mix with $N{=}2000$), and we therefore adopt this setting for all evaluations, including CSJMix and AMI.
\begin{table}[h]
\vspace{-10px}
\centering
\caption{WER (U-WER/B-WER) results on LibriSpeechMix. $\mu$ is the biasing weight at inference time. Best results are bolded.}
\label{tab:biasingweight}
\setlength{\tabcolsep}{2pt}
\renewcommand{\arraystretch}{1.05}
\resizebox{\columnwidth}{!}{%
\begin{tabular}{@{}c c c  cc  cc@{}}
\toprule
 &  &  & \multicolumn{2}{c}{\textbf{LibriSpeech2Mix}} & \multicolumn{2}{c}{\textbf{LibriSpeech3Mix}} \\
\cmidrule(lr){4-5}\cmidrule(lr){6-7}
\textbf{ID} & \textbf{Method} & \textbf{$\boldsymbol{\mu}$} &
$N{=}100\times2$ & $N{=}1000\times2$ & $N{=}100\times3$ & $N{=}1000\times3$ \\
\midrule
\multicolumn{7}{l}{\textbf{Baselines}}\\
\quad B1 & \makecell{A2\\(Table \ref{tab:librispeech_main})} & -- &
4.3 (\textbf{3.3}/12.7) &
4.3 (\textbf{3.3}/12.7) &
9.2 (\textbf{8.3}/17.0) &
9.2 (\textbf{8.3}/17.0) \\
\multicolumn{7}{l}{\textbf{Proposed}}\\
\quad B2 & \multirow{5}{*}{\makecell{A4\\(Table \ref{tab:librispeech_main})}} & 1.0 &
3.7 (3.8/\textbf{2.5}) &
6.6 (6.9/\textbf{4.1}) &
8.6 (9.0/\textbf{4.9}) &
14.4 (15.2/8.2) \\
\quad B3 &  & 0.8 &
3.6 (3.8/2.7) &
6.1 (6.4/\textbf{4.1}) &
8.5 (8.9/5.0) &
13.6 (14.3/8.1) \\
\quad B4 &  & 0.5 &
3.6 (3.7/2.9) &
5.3 (5.4/4.2) &
\textbf{8.3} (8.6/5.3) &
12.0 (12.5/8.0) \\
\quad B5 &  & 0.3 &
\textbf{3.5} (3.6/3.1) &
4.7 (4.8/4.2) &
\textbf{8.3} (8.6/5.8) &
10.7 (11.1/\textbf{7.9}) \\
\quad B6 &  & 0.1 &
3.6 (3.5/4.1) &
\textbf{4.1} (4.0/4.9) &
8.4 (8.5/6.9) &
\textbf{9.1} (9.3/8.3) \\
\bottomrule
\end{tabular}
}
\end{table}
\subsection{CALM: Overall results and analysis}
Table \ref{tab:lsmix_wer_only} shows that CALM (C6/C7) achieves the lowest WERs on LibriSpeechMix, with both reproduced baselines (C4/C5) and prior target-speaker ASR approaches (C1–C3). In our analysis, CALM substantially reduces substitution errors in both all-word and biasing-word categories, indicating enhanced robustness to overlap-induced confusions and more accurate recognition of speaker-relevant terms. Notably, these improvements are attributed to joint acoustic-linguistic modeling, where speaker embeddings directly modulate the dynamic vocabulary layer, aligning acoustic and linguistic information. The auxiliary VAD loss in C7 further reinforces temporal alignment under overlapping speech, achieving the most consistent performance across evaluation conditions.
\vspace{-10px}
\begin{table}[h]
\centering
\caption{WER results on LibriSpeechMix evaluation sets. Best results are bolded. ED represents enrollment duration.}
\label{tab:lsmix_wer_only}
\setlength{\tabcolsep}{3pt}
\renewcommand{\arraystretch}{1.05}
\resizebox{\columnwidth}{!}{%
\begin{tabular}{@{}llcc@{}}
\toprule
\textbf{ID} & \textbf{Method} & \textbf{LibriSpeech2Mix} & \textbf{LibriSpeech3Mix} \\
\midrule
C1 &Whisper-large-SS-TTI \cite{meng24c_interspeech} & 6.99 & 11.40 \\
C2 &Transformer SA-ASR \cite{kanda21b_interspeech} & 6.40 & 8.50 \\
C3 &CONF-TSASR \cite{ConformerTSASR} (ED=7.5s) & 6.30 & 9.00 \\
C4 & A1 (Table \ref{tab:librispeech_main}) (ED=5s) & 4.58 & 9.86 \\
C5 & A2 (Table \ref{tab:librispeech_main}) (ED=5s) \cite{maeda25_interspeech} & 4.28 & 9.21 \\
C6 & A3 (Table \ref{tab:librispeech_main}) (ED=5s) (proposed) & 3.89 & 9.18 \\
C7 & A4 (Table \ref{tab:librispeech_main}) (ED=5s) (Proposed) & \textbf{3.56} & \textbf{8.35} \\
\bottomrule
\end{tabular}%
}
\vspace{-15px}
\end{table}
\subsection{Validation on standardized speech mixtures}
On the AMI-IHM-Mix corpus (Table \ref{tab:ami}), CALM (E4) achieves consistent improvements in B-WER compared to the reproduced baseline (E3), reducing it from 34.7 to 22.1 with a biasing list of $N=1000$. However, unlike in simulated conditions, overall WER increases from 37.4 to 39.1 absolute points. Our error analysis indicates that this degradation is primarily driven by an increase in insertion errors, particularly for short utterances where speaker attribution is more challenging. This effect is most pronounced with smaller list (e.g., $N{=}100$), indicating that the model may misassign these short segments to biasing tokens, resulting in higher WER. We also hypothesize that our use of random enrollment durations to match practical conditions negatively impacts performance compared to \cite{zili1}. Furthermore, as AMI is a conversational dataset, many utterances contain fillers (e.g., \textit{hmmm}, \textit{yeah}), which are rarely included in the biasing list and contribute to effective $N{=}0$ cases. The difference in biasing list construction is also critical: unlike LibriSpeechMix and CSJMix, where lists are built per speaker and concatenated, AMI uses a single global list per segment, altering how overlap is handled and influencing both WER and B-WER. Nevertheless, larger lists appear to mitigate this effect, stabilizing WER and achieving strong B-WER gains, thereby validating CALM’s robustness for biasing term recognition in standardized speech mixtures.
\begin{table}[h]
\vspace{-10px}
\centering
\caption{WER (U-WER/B-WER) results on real AMI-IHM-Mix dataset with enrollment duration (ED). Best results are bolded.}
\label{tab:ami}
\setlength{\tabcolsep}{4pt}
\renewcommand{\arraystretch}{1.05}
\resizebox{\columnwidth}{!}{%
\begin{tabular}{@{}lc cccc@{}}
\toprule
\multirow{2}{*}{\textbf{ID}} & \multirow{2}{*}{\textbf{Method}} & \multicolumn{4}{c}{\textbf{AMI-IHM-Mix}} \\
\cmidrule(lr){3-6}
 &  & $N{=}0$ & $N{=}100$ & $N{=}500$ & $N{=}1000$ \\
\midrule
\multicolumn{6}{l}{\textbf{Baselines}} \\
\quad E1 &\makecell{WavLMBase+ \\ w/ TSE (ED=15s) \cite{zili1}} &
\multicolumn{4}{c}{\makecell{49.5\\(NA/NA)}}  \\
\quad E2 &\makecell{E1 \\ w/ JSM (ED=15s) \cite{zili1}} &
\multicolumn{4}{c}{\makecell{\textbf{28.4}\\(NA/NA)}}  \\
\quad E3 &\makecell{A2 (Table \ref{tab:librispeech_main}) \\ w/ (ED=15s)} &
\multicolumn{4}{c}{\makecell{37.4\\(\textbf{37.7}/34.7)}}  \\
\multicolumn{6}{l}{\textbf{Proposed}} \\
\quad E4 &\makecell{A4 (Table \ref{tab:librispeech_main}) \\ w/ (ED=15s)} &
\makecell{42.3\\(43.2/33.4)} & \makecell{40.3\\(42.0/\textbf{22.5})} & \makecell{39.2\\(40.9/\textbf{22.0})} & \makecell{39.1\\(40.7/\textbf{22.1})} \\
\bottomrule
\end{tabular}%
}
\vspace{-15px}
\end{table}

\subsection{Validation on Japanese dataset}
Table \ref{tab:csjmix_method} presents results on CSJMix, offering a linguistically distinct evaluation from English. CALM achieves consistent gains across both two- and three-speaker mixtures, demonstrating robustness in the Japanese language. In CSJMix2, D2 achieves large reductions in B-CER, e.g., from 16.2 (D1) to 8.3 (D2) on eval1 and from 17.6 (D3) to 8.1 (D4) on eval2 with $N{=}100$, while also lowering overall CER and U-CER. These improvements highlight the effectiveness of combining speaker adaptation with dynamic vocabulary expansion in morphologically richer, character-based languages. In CSJMix3, it continues to reduce B-CER substantially (e.g., from 23.9 in D9 to 11.7 (D10) on eval2 with $N{=}100$), though overall CER remains similar to or slightly above A2 when $N{=}0$ (D7/D9/D11), reflecting the increased difficulty of three-speaker overlap. Overall, CALM maintains stable CER while consistently improving B-CER, validating its generalization across languages.
\vspace{-5px}
\begin{table}[h]
\centering
\small
\setlength{\tabcolsep}{5pt}
\renewcommand{\arraystretch}{1.05}
\caption{CER (U-CER/B-CER) results on CSJMix evaluation sets. Best results are bolded.}
\label{tab:csjmix_method}
\resizebox{\columnwidth}{!}{%
\begin{tabular}{@{}ll l cccc@{}}
\toprule
\multirow{2}{*}{\textbf{ID}} & \multirow{2}{*}{\textbf{Method}} & \multirow{2}{*}{\makecell{\textbf{Eval.} \\ \textbf{Set}}} & \multicolumn{4}{c}{\textbf{Biasing List Size ($N$)}} \\
\cmidrule(l){4-7}
& & & $N{=}0$ & $N{=}100$ & $N{=}500$ & $N{=}1000$ \\
\midrule
\multicolumn{7}{@{}l}{\textbf{CSJMix2} ($N\times 2$)} \\
D1 & A2 (Table \ref{tab:librispeech_main}) & \multirow{2}{*}{eval1} &
\makecell{8.2 \\ (7.2/\textbf{16.2})} &
\makecell{8.2 \\ (7.2/16.2)} &
\makecell{8.2 \\ (7.2/16.2)} &
\makecell{8.2 \\ (\textbf{7.2}/16.2)} \\
D2 & A4 (Table \ref{tab:librispeech_main}) & &
\makecell{\textbf{8.1} \\ (\textbf{7.0}/17.3)} &
\makecell{\textbf{7.1} \\ (\textbf{6.9}/\textbf{8.3})} &
\makecell{\textbf{7.2} \\ (\textbf{7.1}/\textbf{8.1})} &
\makecell{\textbf{7.5} \\ (7.3/\textbf{8.6})} \\
\cdashline{1-7}
D3 & A2 (Table \ref{tab:librispeech_main}) & \multirow{2}{*}{eval2} &
\makecell{7.9 \\ (6.9/17.6)} &
\makecell{7.9 \\ (6.9/17.6)} &
\makecell{7.9 \\ (6.9/17.6)} &
\makecell{7.9 \\ (6.9/17.6)} \\
D4 & A4 (Table \ref{tab:librispeech_main}) & &
\makecell{\textbf{7.2} \\ (\textbf{6.4}/\textbf{16.8})} &
\makecell{\textbf{6.4} \\ (\textbf{6.2}/\textbf{8.1})} &
\makecell{\textbf{6.7} \\ (\textbf{6.5}/\textbf{8.4})} &
\makecell{\textbf{6.9} \\ (\textbf{6.7}/\textbf{8.6})} \\
\cdashline{1-7}
D5 & A2 (Table \ref{tab:librispeech_main}) & \multirow{2}{*}{eval3} &
\makecell{\textbf{8.2} \\ (\textbf{7.1}/\textbf{16.6})} &
\makecell{8.2 \\ (7.1/16.6)} &
\makecell{8.2 \\ (7.1/16.6)} &
\makecell{8.2 \\ (\textbf{7.1}/16.6)} \\
D6 & A4 (Table \ref{tab:librispeech_main}) & &
\makecell{\textbf{8.2} \\ (\textbf{7.1}/17.8)} &
\makecell{\textbf{6.9} \\ (\textbf{6.8}/\textbf{7.7})} &
\makecell{\textbf{7.0} \\ (\textbf{6.9}/\textbf{7.9})} &
\makecell{\textbf{7.4} \\ (7.3/\textbf{8.4})} \\
\midrule
\multicolumn{7}{@{}l}{\textbf{CSJMix3} ($N\times 3$)} \\
D7  & A2 (Table \ref{tab:librispeech_main}) & \multirow{2}{*}{eval1} &
\makecell{\textbf{12.8} \\ (\textbf{11.8}/\textbf{21.2})} &
\makecell{12.8 \\ (\textbf{11.8}/21.2)} &
\makecell{12.8 \\ (\textbf{11.8}/21.2)} &
\makecell{\textbf{12.8} \\ (\textbf{11.8}/21.2)} \\
D8  & A4 (Table \ref{tab:librispeech_main}) & &
\makecell{13.4 \\ (12.4/21.8)} &
\makecell{\textbf{12.1} \\ (12.1/\textbf{11.8})} &
\makecell{\textbf{12.6} \\ (12.6/\textbf{12.4})} &
\makecell{13.6 \\ (13.6/\textbf{13.2})} \\
\cdashline{1-7}
D9  & A2 (Table \ref{tab:librispeech_main})  & \multirow{2}{*}{eval2} &
\makecell{\textbf{12.9} \\ (\textbf{11.7}/23.9)} &
\makecell{12.9 \\ (\textbf{11.7}/23.9)} &
\makecell{12.9 \\ (\textbf{11.7}/23.9)} &
\makecell{\textbf{12.9} \\ (\textbf{11.7}/23.9)} \\
D10 & A4 (Table \ref{tab:librispeech_main}) & &
\makecell{\textbf{12.9} \\ (11.9/\textbf{22.3})} &
\makecell{\textbf{11.9} \\ (11.9/\textbf{11.7})} &
\makecell{\textbf{12.4} \\ (12.4/\textbf{12.6})} &
\makecell{13.2 \\ (13.2/\textbf{13.1})} \\
\cdashline{1-7}
D11 & A2 (Table \ref{tab:librispeech_main}) & \multirow{2}{*}{eval3} &
\makecell{\textbf{11.4} \\ (\textbf{10.3}/\textbf{19.9})} &
\makecell{11.4 \\ (\textbf{10.3}/19.9)} &
\makecell{11.4 \\ (\textbf{10.3}/19.9)} &
\makecell{\textbf{11.4} \\ (\textbf{10.3}/19.9)} \\
D12 & A4 (Table \ref{tab:librispeech_main}) & &
\makecell{11.8 \\ (10.7/21.7)} &
\makecell{\textbf{10.4} \\ (10.4/\textbf{10.6})} &
\makecell{\textbf{10.9} \\ (10.7/\textbf{11.9})} &
\makecell{11.6 \\ (11.5/\textbf{12.3})} \\
\bottomrule
\end{tabular}%
} 
\vspace{-10px}
\end{table}
\vspace{-5px}
\section{Conclusion}
This paper introduced CALM, the first end-to-end framework that integrates target-speaker embeddings with dynamic vocabulary expansion for personalization of multi-speaker ASR. By combining acoustic speaker conditioning with linguistic biasing in a unified architecture, CALM effectively addresses both overlap-induced acoustic errors and unseen vocabulary challenges. Experimental results demonstrate consistent improvements across simulated English (LibriSpeechMix) and Japanese (CSJMix) mixtures, with significant reductions in B-WER. Validation on the standardized speech mixtures (AMI corpus) further highlights robustness under conversational conditions, where fillers and speaker variability pose additional challenges. These findings confirm that acoustic-linguistic integration enables scalable and personalized multi-speaker ASR, advancing beyond the limitations of prior target-speaker and contextual biasing approaches.

\clearpage
\begingroup
\footnotesize
\bibliographystyle{IEEEbib}
\bibliography{icassp2026}
\endgroup

\end{document}